\documentclass[twocolumn]{jpsj3} %% two-column layout
\usepackage{graphicx}
\usepackage{bm}

\title{Antiferromagnetic Order and Superconductivity in Sr$_4$(Mg$_{0.5-x}$Ti$_{0.5+x}$)$_{2}$O$_{6}$Fe$_2$As$_2$ \\ with Electron Doping: $^{75}$As-NMR Study }

\author{Keisuke \textsc{Yamamoto}$^{1}$, Hidekazu \textsc{Mukuda}$^{1,4}$\thanks{E-Mail:mukuda@mp.es.osaka-u.ac.jp}, Hiroaki \textsc{Kinouchi}$^{1}$, Mitsuharu \textsc{Yashima}$^{1,4}$, Yoshio \textsc{Kitaoka}$^{1}$,\\ 
Mamoru Yogi$^{3}$, Shinya \textsc{Sato}$^{2}$, Hiraku \textsc{Ogino}$^{2,4}$, and Jun-ichi \textsc{Shimoyama}$^{2,4}$}

\inst{$^{1}$Graduate School of Engineering Science, Osaka University, Toyonaka, Osaka 560-8531 \\
$^{2}$Department of Applied Chemistry, The University of Tokyo, Hongo, Bunkyo, Tokyo 113-8656\\
$^{3}$Department of Physics and Earth Sciences, Faculty of Science, University of the Ryukyus, Okinawa 903-0213\\
$^{4}$JST, TRIP (Transformative Research-Project on Iron-Pnictides), Chiyoda, Tokyo 102-0075
}

\abst{
We report an $^{75}$As-NMR study on iron (Fe)-based superconductors with thick perovskite-type blocking layers Sr$_4$(Mg$_{0.5-x}$Ti$_{0.5+x}$)$_{2}$O$_{6}$Fe$_2$As$_2$ with $x$=0 and 0.2. We have found that antiferromagnetic (AFM) order takes  place when $x$=0, and superconductivity (SC) emerges below $T_c$=36 K when $x$=0.2. These results reveal that the Fe-pnictides with thick perovskite-type blocks also undergo an evolution from the AFM order to the SC  by doping electron carriers into FeAs planes through the chemical substitution of Ti$^{+4}$ ions for Mg$^{+2}$ ions, analogous to the F-substitution in LaFeAsO compound. 
The reason why the $T_c$=36 K when $x$=0.2 being higher than the optimally electron-doped LaFeAsO with $T_c$=27 K relates to the fact that the local tetrahedron structure of FeAs$_4$ is optimized for the onset of SC. 
}

\recdate{\today}
\kword{Sr$_4$(Mg,Ti)$_{2}$O$_{6}$Fe$_2$As$_2$, perovskite blocks, superconductivity, antiferromagnetism, NMR}

\newpage
\begin{document}
\maketitle

%%========================================================================================================
%\section{Introduction} 

Iron-based high-$T_{\rm c}$ superconductors comprise a two-dimensional layered structure of iron (Fe)-pnictgen ({\itshape Pn}) planes\cite{Kamihara}, which are separated by blocking layers, such as {\itshape Ln}O ({\itshape Ln}=rare earth), alkaline earth metals and alkaline metals, and so on. 
Relatively high superconducting transition has been reported in Fe-pnictides with thick perovskite-type blocking layers, for example, $T_{\rm c}$ is $\sim 47$\,K for Ca$_4$(Mg$_{0.25}$Ti$_{0.75}$)$_3$O$_y$Fe$_2$As$_2$\cite{Ogino2} and $\sim 37$ K for Sr$_4$V$_2$O$_6$Fe$_2$As$_2$(V-42622)\cite{Zhu}, in which an interlayer distance between Fe{\itshape Pn} layers ($L$) is longer than $13$\,\AA\ \cite{Zhu,Ogino2,Sato,Shirage}. 
However, a ground state of undoped (FeAs)$^-$ layer and a relation with the onset of SC in the Fe-pnictides with thick perovskite blocks has not been identified sufficiently, although it is well known that the superconductivity (SC) in Fe-pnictide compounds emerges in close proximity to antiferromagnetism (AFM)~\cite{Kamihara}.
In Sr$_4$(Mg$_{0.5-x}$Ti$_{0.5+x}$)$_{2}$O$_{6}$Fe$_2$As$_2$ (denoted as Mg$_{0.5-x}$Ti$_{0.5+x}$-42622 hereafter), it has been reported that $T_c$ increases from 0 K to 36 K by the substitution of Ti for Mg\cite{Sato}, and  up to 43 K by the application of high pressure~\cite{Kotegawa2}. 
In this compound with $x$=0, the Fe$^{2+}$ state is formally expected as well as other parent Fe-pnictide compounds.\cite{Kamihara}  Therefore, systematic investigations of these compounds will provide us with further insight into the intimate relationship between SC and AFM order inherent in FeAs layers with a highly two-dimensional electronic structure. 

In this Letter, we report $^{75}$As-nuclear magnetic resonance (NMR) study of Mg$_{0.5-x}$Ti$_{0.5+x}$-42622 with $x=$0 and 0.2, which unravels that an AFM order sets in at $x$=0 and an SC state emerges at $x$=0.2 by doping electron carriers into the FaAs layers through the chemical substitution of Ti for Mg.  
We remark that the ground state of (FeAs)$^{-}$ layer of undoped  Mg$_{0.5-x}$Ti$_{0.5+x}$-42622 resembles those of 1111 and 122 systems. 

%%========================================================================================================
%\section{experimental} 

\begin{figure}[tbp]
\centering
\includegraphics[width=8cm]{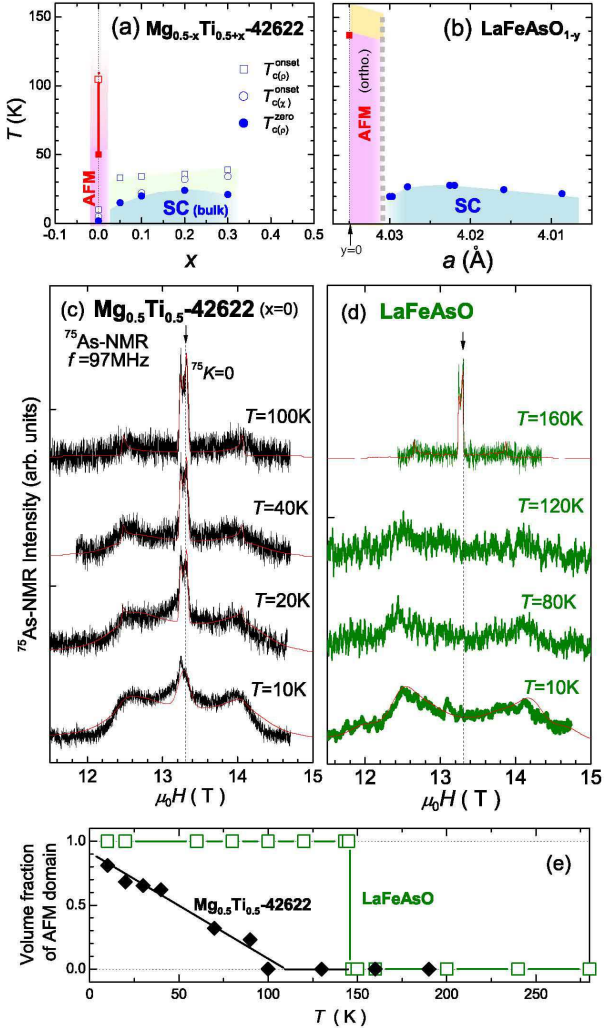}
\caption[]{\footnotesize (Color online) (a) Phase diagram of Mg$_{0.5-x}$Ti$_{0.5+x}$-42622. Respective $T_{c(\rho)}^{\rm onset}$ and $T_{c(\chi)}^{\rm onset}$ are the onset of resistivity drop and SC diamagnetism in  susceptibility, and $T_{c(\rho)}^{\rm zero}$ presents zero resistivity[Sato et al. \cite{Sato}]. The N\'eel temperature ($T_N$) for $x$=0 is $50\sim100$ K, as revealed in this study(see text). (b) Phase diagram of LaFeAsO$_{1-y}$\cite{MukudaFe2}. (c) and (d) are $T$-dependence of $^{75}$As-NMR spectra for $x$=0 and LaFeAsO ($T_N$=140 K), respectively. The solid lines are the simulated spectra when the internal field $H_{int}^{\parallel c}$ and their $\nu_Q$ values at the As site are assumed in these compounds.  (e) $T$ dependence of volume fraction of AFM domain  evaluated from the fractional intensity of the broad spectra. Here, we assume that the difference of spin-spin relaxation time $T_2$ in two phases is neglected.  
}
\label{spec}
\end{figure}

Polycrystalline samples of Mg$_{0.5-x}$Ti$_{0.5+x}$-42622 were synthesized in quartz ampules at ambient pressure as described elsewhere~\cite{Sato}. Powder x-ray diffraction  measurement indicates that the samples are dominated by an intrinsic phase, whereas the $x$=0.2 sample contains small amounts of impurities such as SrFe$_2$As$_2$ and Sr$_2$TiO$_4$~\cite{Sato}. However, the $^{75}$As-NMR signal inherent in Mg$_{0.5-x}$Ti$_{0.5+x}$-42622 at $x$=0.2 is discriminated from that of SrFe$_2$As$_2$ \cite{Kitagawa2}.  Both the samples with $x$=0.0 and 0.2 are the same lattice parameters: $a$- and $c$-axis length of $a$=3.94\AA\ and $c=$15.95\AA, a height of pnictgen from Fe-plane $h_{Pn}\sim~1.4\AA$, and a $Pn$-Fe-$Pn$ bond angle, $\alpha\sim 109.5^\circ$\cite{Ogino2}. These parameters are comparable to the optimum lattice parameters to reach a highest $T_c$ in various series of Fe based compounds, as suggested in the literature~\cite{C.H.Lee,Mizuguchi}. 
The $T_c$s of Mg$_{0.5-x}$Ti$_{0.5+x}$-42622 determined by the resistivity and susceptibility measurements~\cite{Sato} are shown in Fig. \ref{spec}(a). 
Here, $T_{c(\rho)}^{\rm onset}$ and $T_{c(\rho)}^{\rm zero}$ are the respective temperatures for an onset and zero
resistance of SC in resistivity measurement, and $T_{c(\chi)}^{\rm onset}$ is a temperature for an onset of SC diamagnetism in susceptibility measurement. 
Note that $T_{c(\rho)}^{\rm zero}\sim$ 5 K when $x$=0 is not associated with a bulk SC, but a filamentary-induced one, since the SC diamagnetism does not point to a bulk nature \cite{Sato}.
The substitution of Ti for Mg brings about a bulk SC transition at  $T_{c(\rho)}^{\rm onset}\sim$36 K ($T_{c(\rho)}^{\rm zero}\sim$22 K), causing the distinct appearance of SC diamagnetism. 
$^{75}$As-NMR measurements have been performed for coarse powder samples of Mg$_{0.5-x}$Ti$_{0.5+x}$-42622 ($x$=0 and 0.2). 

%fig1------------------------------------------------
%------------------------------------------------
%%========================================================================================================
%\section{results and discussion} 

Figure~\ref{spec}(c) shows the temperature($T$)-dependence of $^{75}$As-NMR ($I$=3/2) spectra for the powder sample of Mg$_{0.5}$Ti$_{0.5}$-42622 with $x$=0. The spectrum at 100 K is a typical powder pattern affected by the nuclear quadrupole interaction in a paramagnetic state. Here, $^{75}\nu_{\rm Q}$ was estimated to be $\sim$11.7 MHz, which is slightly larger than in LaFeAsO.\cite{MukudaNQR} 
As $T$ lowers, the spectrum overlaps with a broad spectrum with its spectral intensity being large below 100 K. 
Note that the broad NMR spectral shape resembles that of the parent compound LaFeAsO, which exhibits the commensurate stripe AFM order below $T_N\sim 140$ K. 
Accordingly, the spectra for Mg$_{0.5}$Ti$_{0.5}$-42622 below 100 K are composed of an AFM ordered phase and a paramagnetic one, which is evidence of phase separation in the sample. 
As shown in Fig. \ref{spec}(e), the volume fraction of the AFM domain evaluated from the fraction of the broad spectra increases at low temperatures, suggesting that the AFM domain size may develop spatially upon cooling in association with a possible inhomogeneity of the local concentration of Mg/Ti atoms.  
As shown in the solid curve in Fig.~\ref{spec}(c), the broad spectra of AFM domains are tentatively reproduced by assuming $^{75}H_{int}^{\parallel c}\sim\pm1.3$ T and $^{75}\nu_{\rm Q}^{\parallel c}$=11.7 MHz. 
On the other hand, it should be noted that the $^{75}$As NMR spectrum of LaFeAsO is well reproduced by assuming an internal field  $^{75}H_{int}^{\parallel c}$=$\pm1.6$ T and $^{75}\nu_{\rm Q}^{\parallel c}$=8.8 MHz, revealing that no phase separation takes place in LaFeAsO, as indicated by the solid line in Fig.~\ref{spec}(d).~\cite{MukudaFe2}
Table~\ref{table1} presents a list of the internal fields at $^{57}$Fe and $^{75}$As sites, $T_N$ and ordered moments $M_{\rm AFM}$ derived from the experiments of $^{57}$Fe-M\"ossbauer and $^{75}$As-NMR on mother compounds of FeAs based  superconductors. The $^{57}H_{int}$ and $^{75}H_{int}$ are induced by $M_{\rm AFM}$ through the respective hyperfine-coupling constants $^{57}A_{hf}$ and $^{75}B_{hf}$. In particular, the origin of $^{75}B_{hf}$ is attributed to an off-diagonal pseudodipole field induced by stripe-type AFM ordered moments lying on the ab-plane at the Fe site~\cite{Kitagawa1}.  Since $^{75}H_{int}^{\parallel c}\sim\pm 1.3$ T in Mg$_{0.5}$Ti$_{0.5}$-42622 is comparable with those values in LaFeAsO and BaFe$_2$As$_2$, it is likely that its AFM ordered state is similar to those in LaFeAsO and BaFe$_2$As$_2$. 
It differs from the static magnetic order of tiny moment from the FeAs layer, which was reported in Sr$_4$Sc$_2$O$_6$Fe$_2$As$_2$(denoted as Sc-42622)\cite{Munevar}. 
Hence, we remark that a commensurate stripe AFM order being comparable to LaFeAsO and BaFe$_2$As$_2$ is realized in the ground state of Mg$_{0.5}$Ti$_{0.5}$-42622, even though M\"ossbauer and neutron scattering experiments in this compound are not yet reported. 

In the powder sample of Mg$_{0.3}$Ti$_{0.7}$-42622 with $x$=0.2, the broad spectrum arising from the AFM domains was not observed, suggesting that doping electron carriers expel the AFM domains. This fact suggests that Ti ions are in a tetravalent state of Ti$^{4+}$ with $3d^0$ in blocking layers, which contrasts with the trivalent state of V$^{3+}$ ions in Sr$_4$V$_2$O$_6$Fe$_2$As$_2$ which are magnetic~\cite{Tatematsu,Kotegawa2,Cao}. Thus, the substitution of nonmagnetic Ti$^{4+}$ ions for Mg$^{2+}$ ions results in an increase in electron density and leads to the collapse of the AFM order. This is also corroborated by the fact that $^{75}\nu_{\rm Q}\sim$12.6 MHz at $x$=0.2 is slightly larger than that at $x$=0, which also resembles the doping dependence of $^{75}\nu_{\rm Q}$ in LaFeAsO system~\cite{MukudaNQR,SKitagawa}.   

%table1  ---------------------------------------------------------
\begin{table}
\centering
\caption[]{\footnotesize  Internal field  at respective $^{57}$Fe and $^{75}$As sites from $^{57}$Fe M\"ossbauer and $^{75}$As-NMR studies for the undoped Fe-pnictides at low temperatures, together with the ordered moment ($M_{\rm AFM}$) and $T_N$\cite{IshidaRev}. 
}

\begin{tabular}{lcccc}
\hline
 & $^{57}H_{\rm int}$[T] & $^{75}H_{\rm int}$[T] & $M_{\rm AFM}$\cite{IshidaRev} & $T_N$[K]\cite{IshidaRev} \\
 & ($^{57}$Fe site) & ($^{75}$As site)     &     &   \\
\hline
Mg$_{0.5}$Ti$_{0.5}$ &              & $\pm$1.3$^\dagger$         &             & 50$\sim$100$^\dagger$ \\
-42622($x$=0) &      &     &   &  \\
LaFeAsO         & 5.3\cite{Kitao}   & $\pm$1.6\cite{MukudaFe2}  & 0.8$\mu_B$\cite{Li}   & 140  \\ 
BaFe$_2$As$_2$  & 5.47\cite{RotterM} & $\pm$1.4\cite{FukazawaM,Kitagawa2} & 0.87$\mu_B$ & 140   \\
SrFe$_2$As$_2$  & 8.9\cite{Tegel}  & $\pm$2.2\cite{Kitagawa2} & 1.01$\mu_B$  &  220  \\
CaFe$_2$As$_2$  & 10\cite{Alzamora} & $\pm$2.6\cite{Beak}      & 0.8$\mu_B$ & 173 \\
Sc-42622 & 1.65\cite{Munevar} &  & 0.11$\mu_B$\cite{Munevar} & 50\cite{Munevar} \\
\hline
\end{tabular}
\label{table1}
\begin{flushleft}
\footnotesize{$\dagger$) results on this experiment}\\
\end{flushleft}
\end{table}
%table1  ---------------------------------------------------------

%fig3------------------------------------------------
\begin{figure}[h]
\centering
\includegraphics[width=8cm]{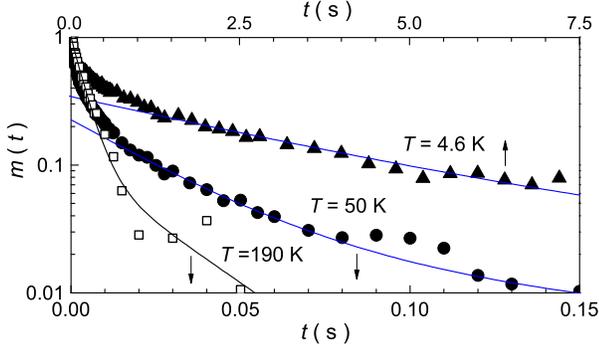}
\caption[]{(Color online) Recovery curves of $^{75}$As nuclear magnetization in $x$=0. The $1/T_1$ can be determined by a single theoretical curve at high temperatures($T>$120 K),  but not below 100 K. The fraction of the long $T_1$ component becomes large at low temperatures in accordance with the emergence of broad spectra below 100 K(See Fig. 1(c)). 
}
\label{fig:recovery}
\end{figure}
%------------------------------------------------

The nuclear spin-lattice relaxation rate $1/T_1$ was measured at the central peak in the $^{75}$As-NMR spectra (see Fig. \ref{spec}(c)). Here, $1/T_1$ was determined from the recovery curve of $^{75}$As nuclear magnetization following the theoretical function for $I=3/2$: $m(t)=[M_0-M(t)]/M_0=0.1\exp(-t/T_1)+0.9\exp(-6t/T_1)$, where $M_0$ and $M(t)$ are the respective nuclear magnetizations for the thermal equilibrium condition and at a time $t$ after the saturation pulse. 
As shown in Fig.~\ref{fig:recovery}, the $m(t)$ at $x$=0 is fitted by a theoretical curve with a single component of $T_1$ at temperatures higher than 120 K, but not lower than $\sim$100 K. Therefore, a long component $T_{1L}$ and a short component $T_{1S}$ are tentatively deduced by assuming an expression given by $m(t)\equiv A_Sm_{S}(t)+A_Lm_{L}(t)$.  Here, $A_S$ and $A_L$ with $A_S+A_L=1$ represent the respective volume fractions of domains with $T_{1S}$ and $T_{1L}$.  Note that $A_L$ becomes larger upon cooling  below $\sim$ 100 K in association with the emergence of AFM domains. Accordingly, it would be expected that $T_{1L}$ and $T_{1S}$ are associated with the AFM domains and the paramagnetic domains, respectively, reflecting the phase separation in Mg$_{0.5}$Ti$_{0.5}$-42622 with $x$=0. 

\begin{figure}[h]
\centering
\includegraphics[width=8cm]{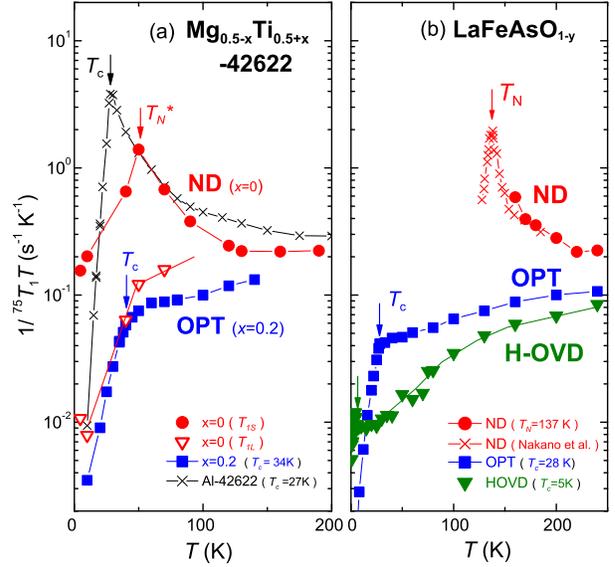}
\caption[]{\footnotesize (Color online) $T$ dependence of $^{75}$As-NMR-(1/$T_1T$) for (a) Mg$_{0.5-x}$Ti$_{0.5+x}$-42622 ($x$=0 and 0.2) and Al-42622 ($T_c=$27 K)\cite{Kinouchi}), and (b) LaFeAsO system (non-doped(ND)\cite{MukudaFe2,Nakano}, optimally-doped(OPT) with $T_c=$28 K\cite{MukudaNQR}, and heavily overdoped(H-OVD) with $T_c=$5 K\cite{Nitta}). }
\label{1/T1T}
\end{figure}

The $T$ dependences of $1/T_{1S}T$ and $1/T_{1L}T$ components are plotted in Fig.~\ref{1/T1T}(a).  
The $1/T_{1S}T$ increases upon cooling below 100 K, but it decreases rapidly with a peak at 50 K, accompanied by a reduction in the volume fraction of the paramagnetic domains. 
On the other hand, $1/T_{1L}T$ decreases gradually upon cooling below $\sim$100 K, accompanied by an increase in the volume fraction of the AFM domains.
These results are consistently interpreted by the fact that AFM ordered domains develop progressively below $\sim$100 K and their fraction exceeds the fraction of paramagnetic domains  below 50 K, as presented in Fig.~\ref{spec}(e).
The peak in $1/T_{1S}T$ at $T_N^*=50$ K may suggest that some paramagnetic  domains undergo an AFM order with a possible distribution of N\'eel temperature ($T_N$) in between 50 K and 100 K, depending on a possible spatial inhomogeneity of the local concentration of Mg/Ti atoms. 
It contrasts with the case of stoichiometric parent compounds LaFeAsO and BaFe$_2$As$_2$, as compared in Fig.~\ref{spec}(e), where the Fe$^{2+}$ states of LaFeAsO and BaFe$_2$As$_2$ are homogeneously realized on the (FeAs)$^-$ layer without any phase separation after the structural transition  to the orthorhombic phase.
We also note that $T_N$s of $x$=0 and Sc-42622 in the previous report\cite{Munevar} are significantly lower than $T_N\sim 140$ K for other parent compounds  LaFeAsO and BaFe$_2$As$_2$, which may relate with the large interlayer distance between the FeAs layers.

%fig4----------------------------------------------------------------------------------------------------
%--------------------------------------------------------------------------------------------------------

Next, we deal with  the $T_1$ results for Mg$_{0.3}$Ti$_{0.7}$-42622 with $x$=0.2, which are shown in Fig.~\ref{1/T1T}(a).
The onset of SC at $T_c$=36 K is also corroborated by a distinct reduction in $1/T_1$.  However, since its broad SC transition prevents us from deducing SC characteristics precisely, we focus only on normal-state properties of this SC compound. We remark that the $T$ dependences of $1/T_1T$ for $x$=0 and 0.2 resemble those in non-doped (ND) and optimally electron-doped LaFeAsO-based compounds~\cite{MukudaFe2,Nakano,MukudaNQR,Nitta}, respectively, as compared in Figs.~\ref{1/T1T}(a) and \ref{1/T1T}(b). The decrease in $1/T_1T$ upon cooling for $x$=0.2 is mostly attributed to the band structure effect~\cite{Ikeda}, suggesting the suppression of AFM spin fluctuations in low energies. 
Since the lattice parameters do not change so much in the series of Mg$_{0.5-x}$Ti$_{0.5+x}$-42622~\cite{Sato},  probably due to the strong covalent bonding in the perovskite blocks, the SC in this compound takes place by increasing Ti$^{4+}$ content through the substitution of Ti for Mg, namely, by doping electron carriers. 
This contrasts with  the optimally electron-doped LaFeAsO with $T_c$=28 K in which either the F-substitution or the O-deficiency changes both the electron-doping level and the lattice parameters.  
In another context, note that the normal-state property for $x$=0 differs from the case of the related 42622 compound Ca$_4$Al$_{2}$O$_{6}$Fe$_2$As$_2$ (denoted as Al-42622)\cite{Shirage}, as shown in Fig.~\ref{1/T1T}(a). The latter compound was characterized by the development of AFM spin fluctuations at low energies\cite{Kinouchi} in association with the nesting between the hole and electron Fermi surfaces (FSs) being quite better\cite{Miyake} owing to the lattice parameters characterized by a very short $a=$ 3.71\,\AA\ , a narrow $\alpha \sim 102.1^\circ$, and a high $h_{Pn} \sim  1.50$\,\AA \cite{Shirage}. 
By contrast, MgTi-42622 with a nearly ideal FeAs$_4$-tetrahedron possesses slightly worse FS nesting properties of than that of Al-42622\cite{Usui}, but $T_c$ is higher than in Al-42622. 
This result suggests that AFM spin fluctuations are not only a unique parameter for enhancing $T_c$.  
We also note that a highest $T_c$=36 K in the series of Mg$_{0.5-x}$Ti$_{0.5+x}$-42622 with the optimum electron doping level at $x$=0.2 is higher than $T_c$=28 K in optimally doped LaFeAsO, which should be ascribed to the fact that the local tetrahedron structure of FeAs$_4$ is optimized. %This suggest that a condition for optimizing SC should be addressed from the lattice structure point of view.
Within a spin-fluctuation mediated pairing theory on a five-orbital model, Usui {\it et al.} have theoretically claimed  that not only the nesting of the hole and electron FSs but also the multiplicity of FSs are important to realize high-$T_c$ SC in Fe based compounds \cite{Usui}. According to this scenario, the higher $T_c$ in Mg$_{0.5-x}$Ti$_{0.5+x}$-42622 can be attributed to the larger multiplicity of FSs in Mg$_{0.5-x}$Ti$_{0.5+x}$-42622 than in Al-42622, whereas the nesting property of FSs are not perfect. 
%In this context, the multiplicity of FSs may be  maximized in Mg$_{0.3}$Ti$_{0.7}$-42622 with $x$=0.2 with the optimum local structure of FeAs tetrahedron, nevertheless the nesting condition is apparently worse than in Al-42622\cite{Usui}, which is one of the possible scenario.  
Further systematic studies on the relationship between the local structure and electronic state in the related 42622 compounds are desired.

%\section{summary} 

In summary, the $^{75}$As-NMR studies on Sr$_4$(Mg$_{0.5-x}$Ti$_{0.5+x}$)$_2$Fe$_2$As$_2$O$_6$ have unraveled that the AFM stripe order takes place for $x$=0 and the SC sets in at $T_c$=36 K for $x$=0.2. The increase of Ti substitution from $x$=0 to 0.2 brings about the onset of SC with $T_c=36$ K as a result of doping electron carriers into FeAs layers, which resembles the variation of the electronic states in the electron-doped LaFeAsO compounds through either F-substitution or O- deficiency.   
The phase diagram of the Fe-pnictides with thick perovskite-type blocking layers resembles those in other Fe-based superconductors which emerge in close proximity to the AFM phase by doping either electron or hole carriers. 
As for the SC state, the comparison with the related 42622 compound Ca$_4$Al$_{2}$O$_{6}$Fe$_2$As$_2$ with $T_c$=27 K suggests that antiferromagnetic spin fluctuations are not a unique factor for enhancing $T_c$. 
The reason why the $T_c$=36 K at $x$=0.2 is higher than the optimally electron-doped LaFeAsO with $T_c$=28 K may relate to the fact that the local tetrahedron structure of FeAs$_4$ is optimized for the onset of SC. 
%, which suggests that the optimum doping of electron carriers does not always lead to the high $T_c$. 
%We propose that a condition for optimizing SC should be addressed from the lattice structure point of view.

%\section*{Acknowledgements}

%{\footnotesize
This work was supported by a Grant-in-Aid for Specially Promoted Research (20001004) and by the Global COE Program (Core Research and Engineering of Advanced Materials-Interdisciplinary Education Center for Materials Science) from the Ministry of Education, Culture, Sports, Science and Technology (MEXT), Japan.
%}

\end{document}